# Phase-selective growth of water-soluble strontium aluminate via sputtering deposition for freestanding oxide membranes

Evgenios Stylianidis[1], Marie Dallocchio[2], Bernat Mundet[3], Lucia Varbaro[4], Clementine Thibault[4], Christo Guguschev[5], Mario Brützam[5], Javier Herrero[6], Marta Gibert[1]

[1] Institute of Solid State Physics, TU Wien, Vienna, Austria

[2] Université Caen Normandie, CEA, ENSICAEN, CNRS, Normandie Univ, CIMAP UMR6252, Caen, France

[3] Catalan Institute of Nanoscience and Nanotechnology (ICN2), CSIC and BIST, Campus UAB, Bellaterra, Spain

[4] Department of Quantum Matter Physics, University of Geneva, Geneva, Switzerland

[5] Leibniz-Institut für Kristallzüchtung, Berlin, Germany

[6] ALBA Synchrotron Light Source, Cerdanyola del Vallès, Spain

## Abstract

We report the epitaxial growth of high-quality water-soluble strontium aluminate films via sputtering deposition. We systematically investigate the dependence of the films structural quality on the growth conditions and we show that we can selectively grow, using the same target, both the cubic $Sr_3Al_2O_6$ and the tetragonal $Sr_4Al_2O_7$ phases by varying the atmosphere during growth. Using these water-soluble sacrificial layers we then synthesize ferromagnetic freestanding $Sm_2NiMnO_6$ membranes. By means of X-ray diffraction, synchrotron spectroscopic techniques and SQUID magnetometry, we demonstrate that our freestanding films exhibit properties comparable to epitaxial films.

## Introduction

The release of epitaxially grown transition metal oxide thin films from the bulk substrate they are grown on, not only offer a vast playground for the exploration of the properties of unclamped, nanoscale single crystalline films [1–3], but also enable the integration of complex oxides onto other platforms for a range of technological applications [4]. In recent years, multiple strategies have been explored for the realization of freestanding membranes. Mechanical lift-off of epitaxial thin films was demonstrated through spontaneous spalling of epitaxial heterostructures [5], through graphene-assisted remote epitaxy [6,7] or with the use of Ni stressor layers [7,8]. Remote epitaxy on top of graphene-buffered substrates preserves epitaxial growth but allows weak Van der Waals forces to develop between the substrate and the grown films, enabling the mechanical peel-off of the films with atomic precision [6]. This method, however, seems to be incompatible with high-energy techniques, such as sputtering, where the harsh plasma damages graphene [9].

The most common and accessible approach for the fabrication of freestanding films is that of the selective chemical etching of epitaxial sacrificial layers. Multiple sacrificial layers have been explored to-date, including ZnO [10] or $La_{0.7}Sr_{0.3}MnO_3$ [11] which require etching in acidic solutions, a process that often damages the freestanding epilayer as well. A decade ago, the epitaxial growth of the perovskite-compatible and water-soluble $Sr_3Al_2O_6$ was first demonstrated [12], setting a new perspective into the research of freestanding transition metal oxide thin films [13]. The ability to release epitaxially grown oxide films from their substrates with common solvents, such as water, offers a much more controllable and accessible pathway for the realization of freestanding nanoscale crystalline oxides.

During the past years, a tremendous progress was made on the epitaxial synthesis of a variety of water-soluble sacrificial layers, covering a wide range of lattice parameters compatible with the vast majority of perovskite, or related structure, transition metal oxides. Whereas cubic $Sr_3Al_2O_6$ has a pseudoperovskite pseudocubic unit cell lattice parameter of 3.96 Å (Figure 1 (a) top) [12], it was later shown that isovalent doping with Ba or Ca can continuously tune the pseudocubic lattice parameter of $(Ba,Sr,Ca)_3Al_2O_6$ solid solutions in a range from 3.82 Å for $Ca_3Al_2O_6$ to 4.13 Å for $Ba_3Al_2O_6$ [14,15]. More recently, a tetragonal $Sr_4Al_2O_7$ phase was discovered [16,17], with a pseudoperovskite pseudocubic lattice parameter of 3.87 Å (Figure 1 (a) bottom) and a remarkable structural flexibility, giving an additional option for lattice compatibility. More recently, the epitaxial growth of the water-soluble hexagonal $BaAl_2O_4$ was reported [18], bringing a new potential for the fabrication of freestanding membranes made out of materials with six-fold or three-fold symmetry.

Yet despite considerable progress in the epitaxial synthesis of high-quality sacrificial layers in recent years, a clear gap remains in how these layers are grown. The vast majority of reports are based on films grown via pulsed laser deposition, molecular beam epitaxy or chemical methods [13,19]. Sputtering, despite being a well-established technique among the physical vapor deposition methods for the epitaxial growth of oxides and the primary technique for many research groups worldwide [20], remains largely underexplored for fully in-situ growth of freestanding oxide membranes. Reports of sputtering-grown strontium aluminate sacrificial layers remain sparse to-date. The group of Xia Hong has reported on freestanding oxide films grown fully in-situ via off-axis radiofrequency (RF) magnetron sputtering, using $Sr_3Al_2O_6$ sacrificial layers grown in highly oxidized conditions with $Ar/O_2$ ratio of 1/8, in an atmosphere of 0.01 mbar at 650°C [21,22], and more recently Pesquera and Santiso reported the sputtering growth of $Ca_3Al_2O_6$ using $Ar/O_2$ ratio of 4/1 in an atmosphere of 0.001 mbar at a temperature of 700°C [13]. To the best of our knowledge, however, a systematic study on the growth conditions of high crystalline quality epitaxial water-soluble strontium aluminate films via sputtering is still lacking.

In this work, we report the growth of high-quality water-soluble strontium aluminate thin films by means of off-axis RF magnetron sputtering. We show that using the same target but varying the atmosphere during growth, we can selectively stabilize the cubic $Sr_3Al_2O_6$ or the tetragonal $Sr_4Al_2O_7$ phases. After establishing the good structural quality of the layers, we demonstrate a good water solubility and produce ferromagnetic $Sm_2NiMnO_6$ freestanding membranes transferred on Si support substrates with properties comparable to those of epitaxial films.

## Results and discussion

We used a home-built off-axis RF magnetron sputtering system equipped with an in-situ RHEED and a loadlock insertion system [23]. For the growth of strontium aluminate thin films, we used a 2-inch nominal $Sr_4Al_2O_7$ ceramic target. In Figure 1 (b-d), we present representative RHEED and X-ray diffraction results for two sputtering-grown strontium aluminate films on (001)-oriented $SrTiO_3$ substrates using different growth conditions, which were optimized for each phase. We note that all strontium aluminate films shown are capped with a 5 nm-thick layer of $Sm_2NiMnO_6$. As evidenced by our results, by varying the growth conditions, we can stabilize the two known distinct phases of strontium aluminate from the same target: the cubic $Sr_3Al_2O_6$ and the tetragonal $Sr_4Al_2O_7$. RHEED oscillations are a testimony of a layer-by-layer growth of the two films. Furthermore, the dynamic RHEED signal shows additional periodicities which are related to the structure of each phase. For the case of $Sr_3Al_2O_6$ we observe bundles of two oscillations repeating. Since the cubic $Sr_3Al_2O_6$ unit cell consists of four individual sublayers along the [001] direction, each two-oscillation bundle reflects the growth of half cubic unit cell [24]. For $Sr_4Al_2O_7$, bundles of three oscillations are observed. Because the tetragonal unit cell of $Sr_4Al_2O_7$ has six individual sublayers along the [001] direction, each three-oscillation bundle shows the growth of half tetragonal $Sr_4Al_2O_7$ unit cell [17]. The oscillations periodicity is a distinct sign of the grown phase and reflects the long-range structural coherence of the grown layer [24]. X-ray diffraction (Figure 1 (d)) around the 001 Bragg reflection of $SrTiO_3$ indeed confirms the stabilization of the two different phases: $Sr_3Al_2O_6$ $004_c$ Bragg peak is observed at around 22° with a lattice parameter of 15.724 Å, while $Sr_4Al_2O_7$, with its $006_t$ Bragg peak located at around 21°, has a lattice parameter of 25.33 Å. Laue oscillations in the X-ray diffractograms show the high structural quality of the sputtering-grown films. We note that the Bragg peaks of the 5 nm-thick $Sm_2NiMnO_6$ layers are not discernible as they coincide with the substrate peaks.

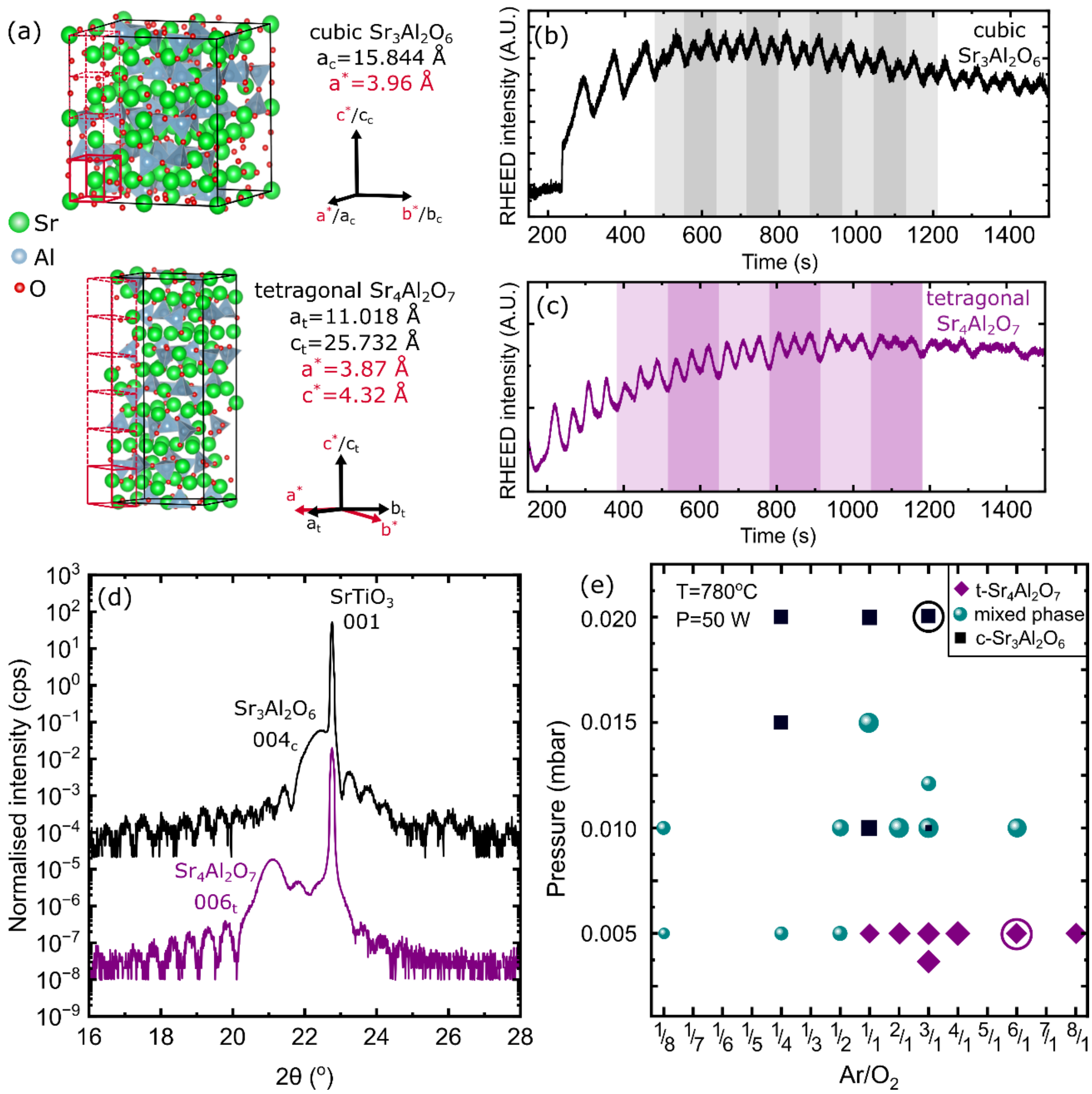

**Figure 1**: Selective-growth of cubic $Sr_3Al_2O_6$ and tetragonal $Sr_4Al_2O_7$ films by sputtering deposition. **(a)** Cubic unit cell of $Sr_3Al_2O_6$ (top, black cage) and tetragonal unit cell of $Sr_4Al_2O_7$ (bottom, black cage). The pseudoperovskite unit cells are depicted with red outline. The asterisks denote the pseudoperovskite lattice parameters and principal crystallographic axis. **(b)** and **(c)** Specular-spot RHEED intensity as a function of time during the growth of cubic $Sr_3Al_2O_6$ and tetragonal $Sr_4Al_2O_7$ by sputtering, respectively. The shadings in each figure depict the periodicity of the RHEED signal, as described in the main text. **(d)** X-ray diffraction of the two samples shown in (b) and (c) around the 001 Bragg peak of the $SrTiO_3$ substrate. The $Sr_3Al_2O_6$ and $Sr_4Al_2O_7$ films are 16 nm and 18 nm, respectively. The films are capped with 5 nm-thick $Sm_2NiMnO_6$ layers. **(e)** Pressure versus $Ar/O_2$ ratio phase diagram for sputtering-grown strontium aluminate thin films. The markers are to scale with the thickness of the respective films. The black and magenta circles indicate the growth conditions of the films presented in Figure 1 (b-d) and Figure 2 (a).

Figure 1 (e) presents our experimentally determined phase map for sputtering-grown strontium aluminate films as a function of the pressure and $Ar/O_2$ ratio, grown at 780°C and 50 W RF power. A more detailed presentation of the growth conditions study can be found in Supplementary Section A. Overall, the phase diagram of Figure 1 (e) shows that the two phases can be stabilized using the same ceramic target at different pressure regimes, as previously observed with films grown via PLD [16]. The cubic $Sr_3Al_2O_6$ phase is stabilized at relatively higher pressures (around 0.02 mbar), while the tetragonal $Sr_4Al_2O_7$ phase requires lower pressures (around 0.005 mbar). At intermediate pressures, mixed-phase films are obtained (see X-ray diffractograms in Supplementary Figure A2). Furthermore, the cubic $Sr_3Al_2O_6$ phase can be stabilized in a wide range of gaseous $Ar/O_2$ ratios, including $O_2$-rich conditions in agreement with previously sputtering-grown $Sr_3Al_2O_6$ films [21], while an $O_2$-rich environment is not favourable for the tetragonal $Sr_4Al_2O_7$ phase. The reproducibility of the obtained conditions was tested in a different sputtering system using a nominal $Sr_3Al_2O_6$ ceramic target, and as depicted in Supplementary Figure B1, high-quality $Sr_3Al_2O_6$ films were obtained. Parenthetically, we note that the reproducibility of the growth conditions is linked to the condition of the ceramic target. Although our systems are equipped with loadlock insertion systems and the main sputtering chambers are not exposed to atmosphere, we found that the surface of the ceramic targets showed signs of degradation (change in color) after remaining unused for several weeks, and a prolonged pre-sputtering was necessary to recover the growth conditions.

In Figure 2, we present detailed structural characterization for optimally grown strontium aluminate films (growth conditions are indicated with the black and magenta circles in Figure 1 (e)), capped with a 5 nm-thick $Sm_2NiMnO_6$ layer: a 17 nm-thick $Sr_4Al_2O_7$ film grown on $(110)_o$-oriented $NdGaO_3$ substrate ($a_{pc}$=3.863 Å) with a lattice mismatch of $\varepsilon = -0.7\%$, and a 19 nm-thick $Sr_3Al_2O_6$ film grown on (001)-oriented $SrTiO_3$ substrate (a=3.905 Å) with a lattice mismatch of $\varepsilon = -1.4\%$. The specular X-ray diffractograms (Figure 2 (a)) are evidence of the high-quality phase-pure strontium aluminate films. The finite-thickness Laue oscillations indicate the long-range crystalline coherence of the films. Phase identification for the two films is achieved by comparing the experimental diffraction data with the simulated specular powder diffraction (lower panel in Figure 2 (a)), showing an excellent agreement. We highlight the presence of the $0010_t$ peak of $Sr_4Al_2O_7$, which is very sensitive to structural distortions in the film, occurred for example by moisture degradation.

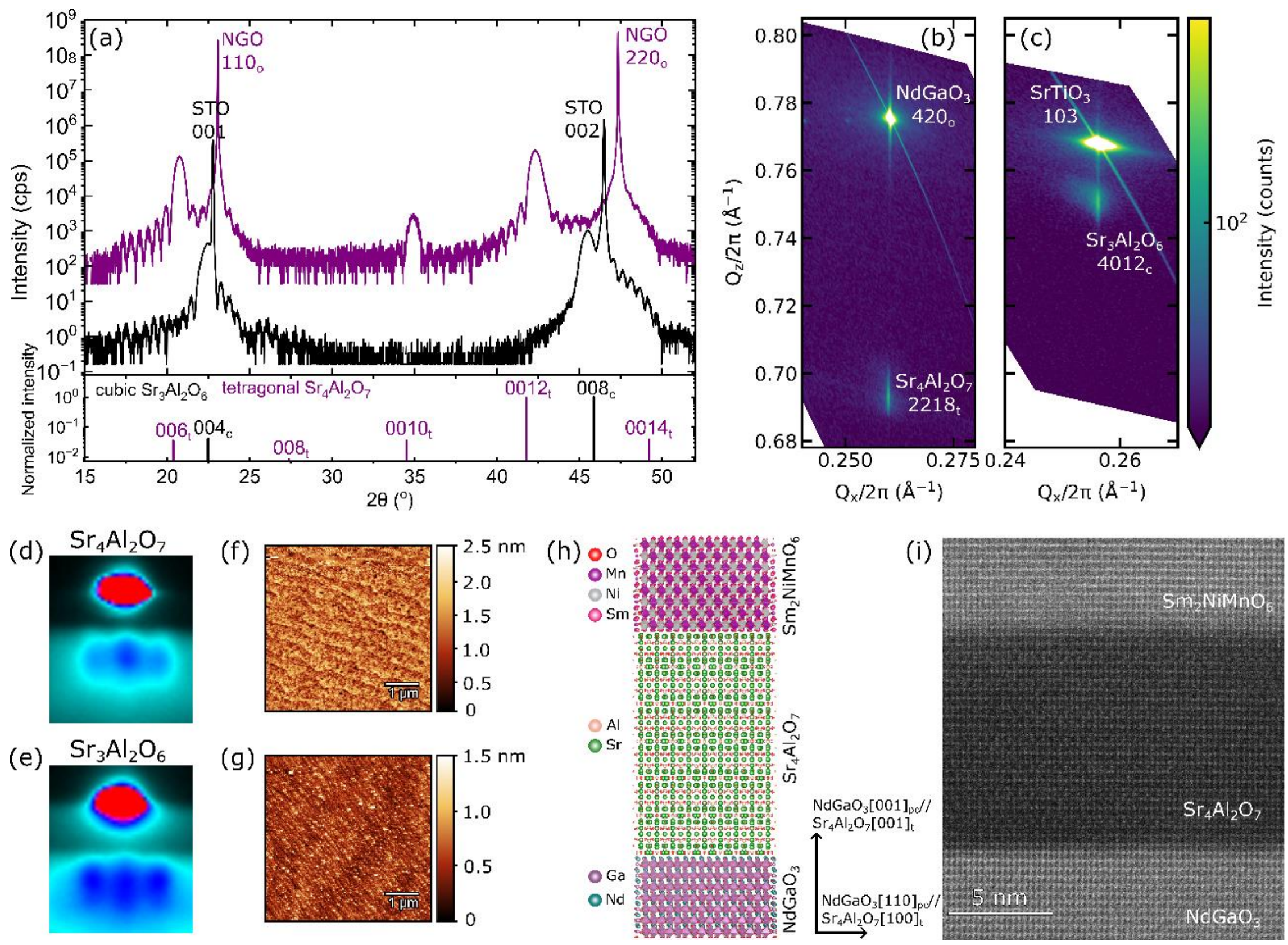

**Figure 2**: Structural characterization of the sputtering-grown strontium aluminate films. **(a)** Specular X-ray diffractograms for a 17 nm-thick $Sr_4Al_2O_7$ film grown on $(110)_o$-oriented $NdGaO_3$ (red) and a 19 nm-thick $Sr_3Al_2O_6$ film grown on (001)-oriented $SrTiO_3$. The growth conditions for these films are shown with magenta and black circles in the phase diagram of Figure 1 (e). Both films are capped with 5 nm $Sm_2NiMnO_6$. Lower panel: specular 00L powder diffraction for the cubic $Sr_3Al_2O_6$ and the tetragonal $Sr_4Al_2O_7$. **(b)** and **(c)** Reciprocal space maps around the off-specular $420_o$ reflection of $NdGaO_3$ and 103 reflection of $SrTiO_3$, respectively, for the samples shown in (a). **(d)** and **(e)** RHEED images for a $Sr_4Al_2O_7$ and a $Sr_3Al_2O_6$ film, respectively. The images were recorded at a $[100]_{pc}$ incident electron beam direction. **(f)** and **(g)** AFM images of the $Sm_2NiMnO_6$ capping layers for the sample with tetragonal $Sr_4Al_2O_7$ and cubic $Sr_3Al_2O_6$, respectively. **(h)** Cross-sectional schematic with atomic resolution of a heterostructure consisting of a $Sr_4Al_2O_7$ layer grown on $[001]_{pc}$ $NdGaO_3$ substrate with a $Sm_2NiMnO_6$ capping layer, viewed along the $[110]_{pc}$ zone axis. **(i)** Cross-sectional HAADF-STEM image of a similar heterostructure imaged along the $[110]_{pc}$ zone axis.

Reciprocal space map (RSM) around off-specular Bragg reflections of the substrates are presented in Figure 2 (b) and (c). The $Sr_4Al_2O_7$ film appears strained on the $NdGaO_3$ substrate, while the cubic $Sr_3Al_2O_6$ shows slight structural relaxation with respect to the $SrTiO_3$ substrate. Furthermore, we mapped the off-specular reflections of $Sr_4Al_2O_7$ at different azimuths (Supplementary Figure C1) and we found that the $Sr_4Al_2O_7$ film retains its tetragonal symmetry, despite the orthorhombic symmetry of the $NdGaO_3$ substrate. RHEED images of the strontium aluminate films prior to $Sm_2NiMnO_6$ growth show stripe-like patterns (Figure 2 (d) and (e)), characteristic of smooth surfaces. AFM images atop the $Sm_2NiMnO_6$ capping layers indicate smooth surfaces with discernible steps and terraces due to the substrates miscut (Figure 2 (f) and (g)). We also note that we have successfully grown high-quality $Sr_4Al_2O_7$ films on various single crystalline substrates, as shown in Supplementary Figure C2.

To gain atomic information on the structure of the $Sr_4Al_2O_7$ films we performed cross-sectional aberration-corrected scanning transmission electron microscopy (STEM) measurements as shown in Figure 2 (i). The acquired high-angle annular dark-field (HAADF-STEM) image, taken along the $[110]_{pc}$ zone axis shows sharp interfaces between the different layers. The atomic arrangement in the $Sr_4Al_2O_7$ layer shows great resemblance to the expected tetragonal structure (schematic in Figure 2 (h)) confirming the high-quality of our films. We note that small areas resembling the cubic $Sr_3Al_2O_6$ structure were observed. However, due to the great sensitivity of the film to the electron beam which causes instant amorphization upon exposure (Supplementary Figure C3), it is difficult to conclude whether traces of the cubic phases were present in the as-grown film or if they were formed during imaging.

Next, we examined the water solubility of the sputtering-grown strontium aluminate films. As shown in Figure 3, both $Sr_4Al_2O_7$ and $Sr_3Al_2O_6$ are highly soluble in high-purity deionized (DI) water, albeit with different solubility speeds. As shown in the optical microscope images of Figure 3 (a), the 17 nm tetragonal $Sr_4Al_2O_7$ film was fully dissolved within 20 minutes in DI water, while the 17 nm cubic $Sr_3Al_2O_6$ film was fully dissolved after 3 hours. The difference in solubility between the two phases is in agreement with films grown via pulsed laser deposition [16]. We also found that the solubility of the films is highly dependent on the thickness of the strontium aluminate layer. We found that $Sr_4Al_2O_7$ films thinner than 10 nm show no signs of dissolution even after 8 hours in DI water.

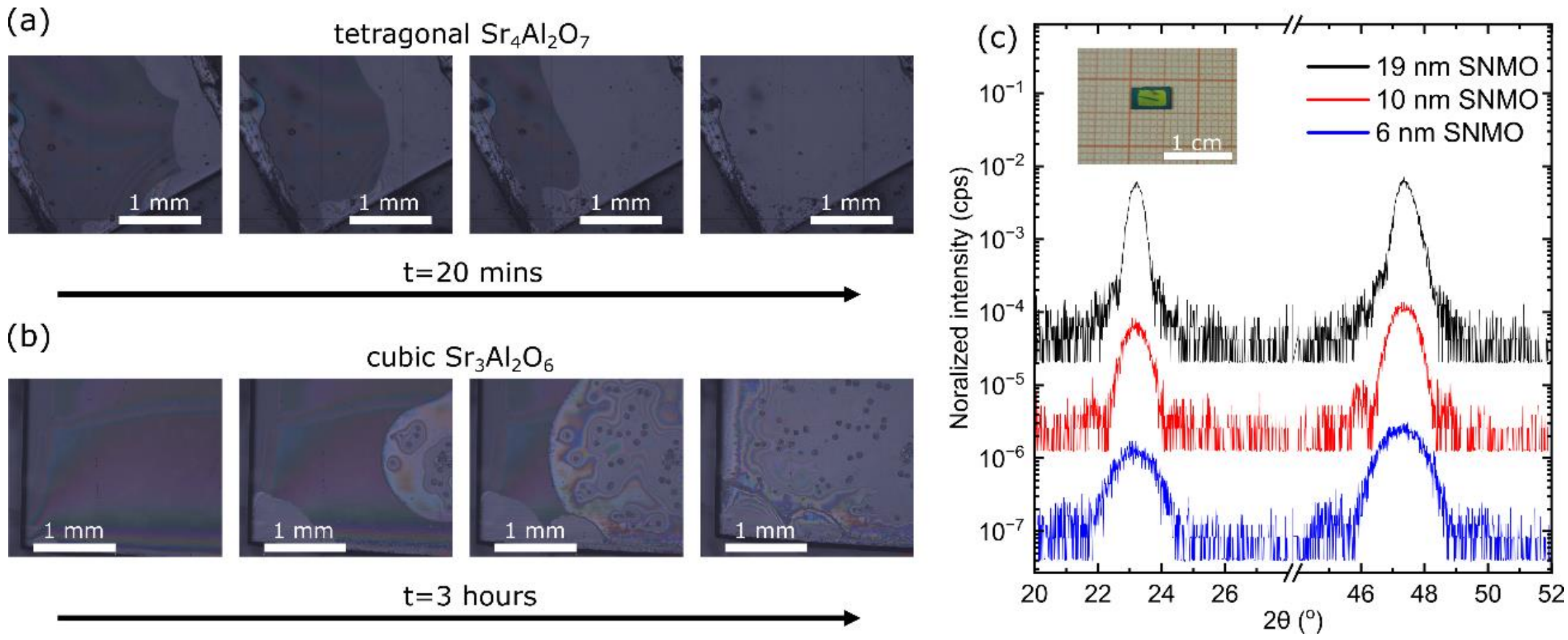

**Figure 3:** Hydrolysis and transfer of freestanding $Sm_2NiMnO_6$ membranes. **(a)** and **(b)** Optical microscope images during the hydrolysis process of tetragonal $Sr_4Al_2O_7$ and cubic $Sr_3Al_2O_6$ sacrificial layers, respectively, capped with 5 nm-thick $Sm_2NiMnO_6$ and spin-coated with PMMA. **(c)** Specular X-ray diffraction measurements for freestanding $Sm_2NiMnO_6$ films of various thicknesses transferred on Si. Inset: optical microscope image of a transferred $Sm_2NiMnO_6$ membrane (yellowish color) on Si substrate.

To demonstrate the use of the sacrificial layers for the production of freestanding oxide layers, we fabricated freestanding $Sm_2NiMnO_6$ membranes. For the best lattice matching, the tetragonal $Sr_4Al_2O_7$ sacrificial layer was used and the heterostructure was grown on a (001)-oriented NSAT substrate ($Nd_{0.396}Sr_{0.604}Al_{0.698}Ta_{0.302}O_3$, a=3.84 Å), which has a lattice mismatch of $\varepsilon = 1.3\%$ with $Sr_4Al_2O_7$, and $\varepsilon = 0.17\%$ with $Sm_2NiMnO_6$ ($a_{pc}$=3.836 Å). After the dissolution of $Sr_4Al_2O_7$, we transferred mm-scale $Sm_2NiMnO_6$ membranes of various thicknesses onto $Si/SiO_2$ support substrates, using the process described in the Methods section. In Figure 3 (c) we show X-ray diffraction measurements for representative freestanding $Sm_2NiMnO_6$ films. The diffractograms show subtle Laue oscillations around the film peaks suggesting that the long-range structural coherence is conserved after the transfer process. We note, however, that the transferred films show locally cracks and wrinkles, and the transfer process requires further optimization (Supplementary section D).

We then turn our attention to the properties of the transferred membranes. First, we examine the electronic structure of the $Sm_2NiMnO_6$ films grown on top of the strontium aluminate sacrificial layers. The double perovskites $RE_2NiMnO_6$ are an intriguing class of transition metal oxides that exhibit ferromagnetism while being electrically insulating. The structure of these materials is similar to a perovskite structure but with two inequivalent B-sites that alternate in all primary crystallographic directions, occupied by Ni and Mn. Charge transfer between Ni and Mn leads to $Ni^{2+}$ and $Mn^{4+}$ oxidation states, and ferromagnetism in these compounds arises from the positive superexchange interaction between $Ni^{2+}$ and $Mn^{4+}$ [25]. In Figure 4 (a-d) we present a simultaneously acquired HAADF-STEM image and energy-dispersive X-ray spectroscopy (EDS) maps for Ni and Mn taken along the $[110]_{pc}$ zone axis from a $Sm_2NiMnO_6$ film grown on a $Sr_4Al_2O_7$ sacrificial layer, confirming the alternate ordering of Ni and Mn. To get information about the valence state of the ions we performed synchrotron X-ray absorption spectroscopy (XAS) around the Ni and Mn $L_{3,2}$ absorption edges of a 5 nm-thick $Sm_2NiMnO_6$ freestanding film transferred on Si compared to an epitaxial 10 nm-thick $Sm_2NiMnO_6$ film grown on a $SrTiO_3$ substrate, as shown in Figure 4 (e) and (f). Comparing with reference spectra [23,26], we conclude a pure $Ni^{2+}$ and $Mn^{4+}$ oxidation states in both epitaxial and freestanding films. These results indicate that the freestanding $Sm_2NiMnO_6$ films retain a long-range ordering of the $Ni^{2+}$-$Mn^{4+}$ sublattice [27,28] even after the transfer process.

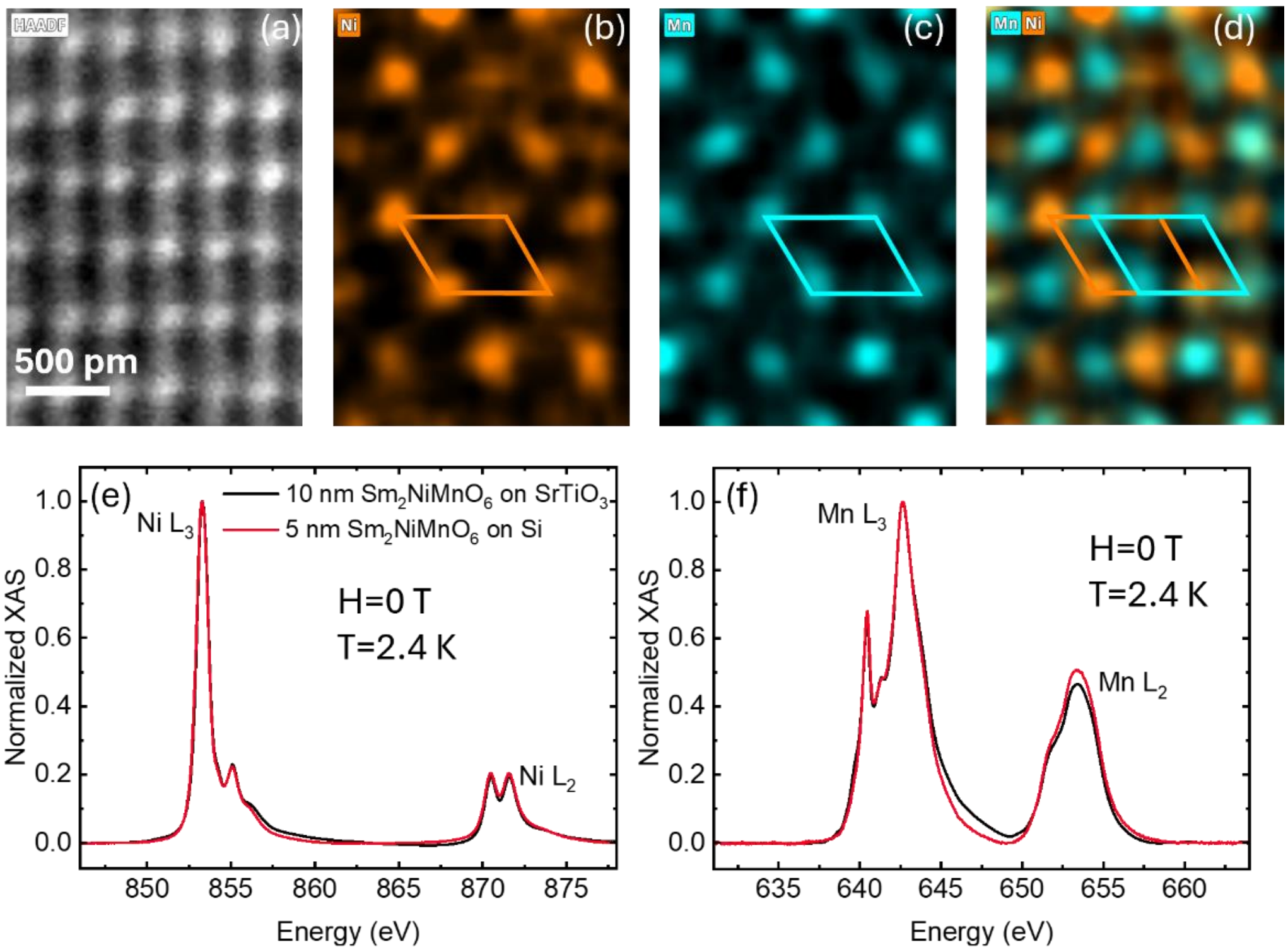

**Figure 4**: Ni-Mn ordering and electronic structure. **(a)** HAADF-STEM image of a $Sm_2NiMnO_6$ film grown on $Sr_4Al_2O_7$ buffered $NdGaO_3$ substrate, imaged along a $[110]_{pc}$ zone axis. **(b-d)** EDS false-color maps for Ni, Mn and mixed Ni-Mn signal, respectively. **(e)** and **(f)** XAS spectra around

the Ni and Mn $L_{3,2}$ edges, respectively, for a 5 nm-thick $Sm_2NiMnO_6$ freestanding film transferred on Si (red), and a 10 nm-thick $Sm_2NiMnO_6$ epitaxial film on $SrTiO_3$ (black).

After establishing the right electronic order in $Sm_2NiMnO_6$, we turn to macroscopic magnetic characterization by means of SQUID magnetometry. In Figure 5 (a), we show the field-cooled magnetization (in $\mu_B$/f.u. where f.u. stands for a formula unit equal to the double perovskite unit cell) as a function of temperature measured with a 0.5 T field for a 15 nm-thick epitaxial $Sm_2NiMnO_6$ film on $SrTiO_3$ substrate and a 19 nm-thick freestanding $Sm_2NiMnO_6$ transferred on $Si/SiO_2$. Both films exhibit a paramagnetic-to-ferromagnetic transition with a Curie temperature near 170 K, close to the bulk transition temperature of $Sm_2NiMnO_6$ [29], and a magnetization downturn near 15 K, which is generally attributed to the magnetically active $Sm^{3+}$ ion [29]. We further performed magnetization versus field measurements at 1.8 K for the two samples as shown in Figure 5 (b). Both samples show a ferromagnetic behavior with similar coercivity and saturation magnetization around 4.45 $\mu_B$/f.u., close to the expected nominal value of bulk specimens [29]. We note the overall lower magnetization of the freestanding film compared to the epitaxial layer at low magnetic fields, which may be a result of the cracks and wrinkles and deserves further investigation.

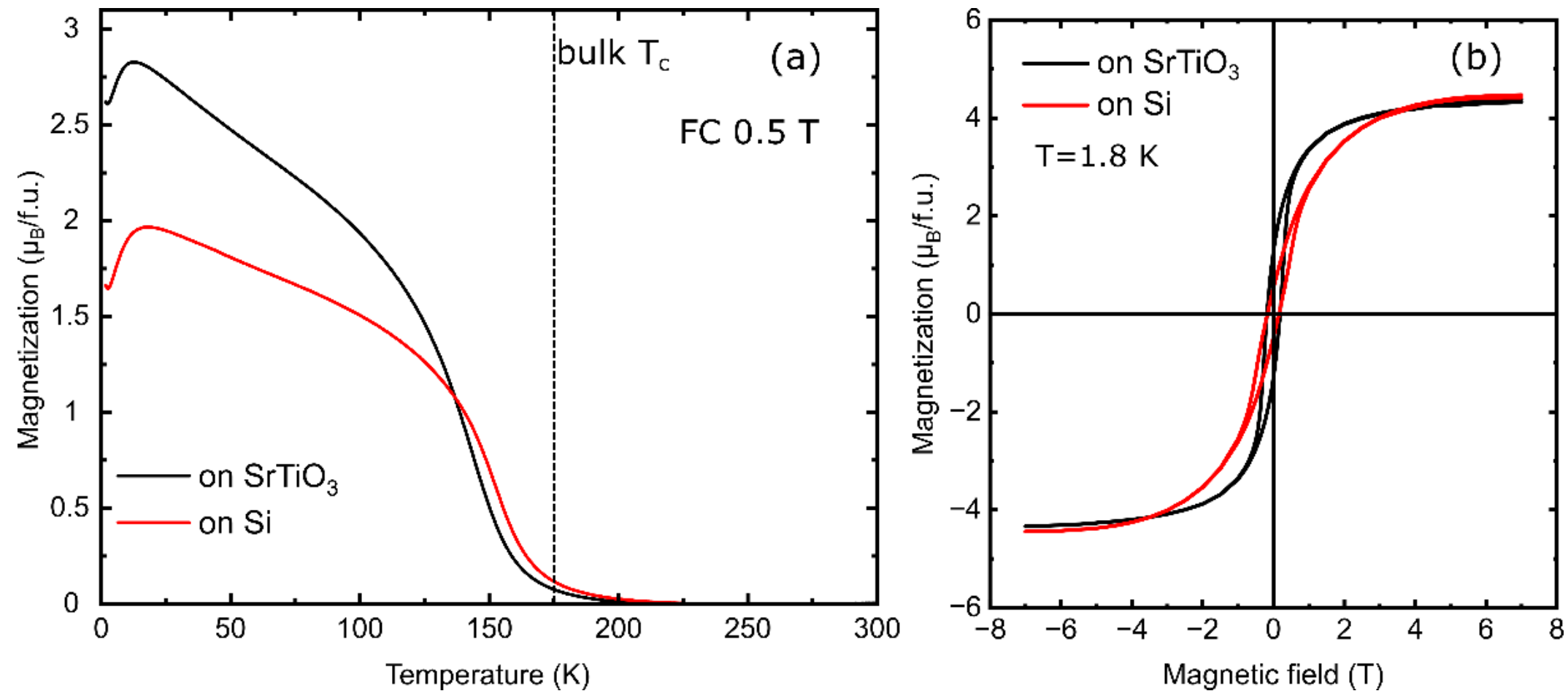

**Figure 5**: Magnetic properties of epitaxial and freestanding $Sm_2NiMnO_6$ thin films. **(a)** Field-cooling (FC) magnetization as a function of temperature measured via SQUID magnetometry with a 0.5 T external field for a 15 nm-thick film epitaxial on $SrTiO_3$ substrate (black) and a 19 nm-thick freestanding film transferred on $Si/SiO_2$ (red) (f.u. stands for a formula unit equal to the double perovskite unit cell). **(b)** Magnetization vs field loops at 1.8 K for the epitaxial and freestanding films.

## Conclusions

We have reported a detailed study of the sputtering growth of epitaxial, high crystalline quality, water-soluble strontium aluminate thin films. We have shown in detail that by varying the growth conditions we were able to stabilize both the tetragonal $Sr_4Al_2O_7$ and the cubic $Sr_3Al_2O_6$ phases using the same ceramic target with great structural quality and water solubility. Using the optimized films, we have showcased the successful fabrication of ferromagnetic double perovskite $Sm_2NiMnO_6$ membranes, with properties similar to epitaxial films. Our study forms a robust foundation towards fully in-situ fabrication of freestanding oxides using sputtering deposition.

## Experimental methods

Growth, dissolution and transfer process

The materials in this work were grown by off-axis RF magnetron sputtering, equipped with a loadlock insertion system and a reflection high-energy electron diffraction (RHEED) setup for the in-situ probe of the growth process. The growth conditions for the strontium aluminate films can be found in Figure 1 (e) of the main text. The $Sm_2NiMnO_6$ films were grown with $Ar/O_2$ ratio of 35/10, atmosphere pressure of 0.1 mbar, RF power at 35 W and a growth temperature of 680°C.

The NSAT substrates with a lattice parameter of about 3.84 Å used this study were grown using a Czochralski method from melt of composition $Nd_{0.396}Sr_{0.604}Al_{0.698}Ta_{0.302}O_3$ or equivalently $(NdAlO_3)_{0.396}$–$(SrAl_{0.5}Ta_{0.5}O_3)_{0.604}$ [30].

For the transfer process, the samples were first spin-coated with poly(methyl methacrylate) (PMMA, with a concentration of 950 kg/mol) at a speed of 4000 rps for 60 s and post-baked at 135°C for 7 minutes. Then, the edges of the samples were polished to facilitate the dissolution process. A piece of PDMS (polydimethylsiloxane) was then placed on top of the samples for additional mechanical support, and the whole system was mounted on a glass slide. The system was then immersed in deionized mili-Q water until the sacrificial layer was fully dissolved and the substrate was separated. Using a transfer stage purchased from hq graphene, the glass-slide/PDMS/PMMA/membrane system was then placed and pressed for 10 minutes on top of a piece of Si, coated with 300 nm of $SiO_2$, which was heated at 100°C. In this stage, the PDMS becomes less adhesive and leaves the PMMA/membrane on the silicon substrate. For the removal of PMMA, the samples were heated on a hot plate at 420°C for 10 to 20 minutes, until the PMMA decomposes thermally, followed by a standard cleaning procedure with acetone and isopropanol.

Structural characterization (X-rays and AFM)

X-rays characterization of the specimens was performed in the X-ray center of TU Wien with a Panalytical Empyrean and a Rigaku Smartlab diffractometers, both equipped with a Ge (220) monochromator using Cu $K\alpha_1$ radiation with wavelength of λ=1.5406 Å. Atomic force microscopy (AFM) imaging was performed using a Park NX10 microscope.

Scanning transmission electron microscopy

High-angle annular dark-field (HAADF) images were acquired using a double aberration corrected scanning transmission electron microscope (STEM) on a Thermofisher SPECTRA 300 operated at 300 kV, using a convergence semi-angle of 20 mrad. Because of the sample reactivity under the electron beam, a set of 12 images were acquired using a probe current of around 5 pA, which were then drift corrected using a rigid registration.

EDS compositional maps were acquired using a 4 quadrant super-X detector and the generated maps were obtained following the Cliff-Lorimer approximation.

SQUID magnetometry

Characterization of the net magnetic properties of the samples was performed using SQUID (superconducting quantum interference device) magnetometry in a Quantum Design MPMS3 system in a vibrating sample magnetometry (VSM) measurement mode. Substrate contributions were removed by premeasuring magnetization as a function of temperature before the film growth ($SrTiO_3$) or the transfer of the freestanding membrane ($Si/SiO_2$).

X-ray absorption spectroscopy

XAS experiments were performed at the BOREAS beamline of the ALBA synchrotron light source in Barcelona, Spain [31]. The measurements were acquired at a grazing incidence geometry, at an angle of 20° with the sample's normal, using the total electron yield detection mode. The XAS data at 0 T are the average of two scans with circularly polarized light.

## Acknowledgments

We acknowledge Lingfei Wang and Liang Si for useful discussions. We would also like to thank Saptam Ganguly, Yaqi Li and Greta Segantini for their valuable insights into the lift-off and transfer process of the freestanding films and the personnel of the X-ray center at TU Wien for technical support. Authors acknowledge the use of instrumentation as well as the technical advice provided by the Joint Electron Microscopy Center at ALBA (JEMCA), carried out under the proposal number 20260400361, and funding from Grant IU16-014206 (METCAM-FIB) to ICN2 funded by the European Union through the European Regional Development Fund (ERDF), with the support of the Ministry of Research and Universities, Generalitat de Catalunya. Electron microscopy experiments were performed at ALBA EM02. XAS experiments were performed in ALBA BOREAS under proposal 20250370406. We are also grateful to Fangyuan Yang for the use of optical microscope and the group of "Physics of three-dimensional nanomaterials" at TU Wien for the use of the spin-coater.